\documentclass[11pt]{article}
\usepackage{graphicx,color} 
\usepackage{amsmath,mathrsfs,amssymb}
\usepackage{upgreek}
\usepackage{natbib}

\usepackage[english]{babel}
\usepackage[T1]{fontenc}

\renewcommand{\theenumi}{(\kern -0.15ex{\roman{enumi}})}

\renewcommand{\th}{${}^\mathrm{th}$ }

\begin{document}
\onecolumn

\noindent \textbf{\LARGE Identity and difference: how topology
 helps to understand  quantum indiscernability} \\[1cm]
Amaury Mouchet \\[2cm]
\let\thefootnote\relax\footnote{Amaury Mouchet\\
Institut Denis Poisson de Math\'ematiques et de Physique Th\'eorique, Universit\'e  Fran\c{c}ois Rabelais de Tours\\ 37200 Tours, France}

\newcommand{\thefootnote}{\arabic{footnote}}
\setcounter{footnote}{0}

\vspace{-2\baselineskip}

\begin{flushright}
  \textit{Soy esos otros,\\
tambi\'en. La eternidad est\'a en las cosas\\ del tiempo, que son formas presurosas\footnote{I am those others too. The eternity is in the things of time, which are precipitate
    forms (trad. \textsc{am}).}.}\\
  Jorge Luis Borges. \emph{Al hijo} (1967) in \emph{El otro, el mismo.}
  \\
  
\end{flushright}
\begin{abstract} This contribution, to be published in Imagine Math 8 to celebrate Michele Emmer's 75th birthday,
  can be seen as the second part of my previous considerations on the relationships between topology and physics \citep{Mouchet18a}.
  Nevertheless, the present work  can be read independently. The following mainly focusses on the connection
  between topology and quantum statistics. I will try to explain to the non specialist how
  Feynman's interpretation of quantum processes through interference
  of classical paths (path integrals formulation),  makes the dichotomy  between
  bosons and fermions quite natural in three spatial dimensions. In (effective) two dimensions,
  the recent experimental evidence 
  of intermediate statistics (anyons)\citep{Bartolomei+20a} comfort that  topology (of the braids) provides a fertile soil for our understanding
  of quantum particles.  
\end{abstract}
\nocite{Emmer/Abate18a}

\section{Old puzzles}

Among the most common primary concepts
that are jeopardised by quantum physics, the related notions of identity, individuality or discernability are not the
least\footnote{See for instance the contributions in \citep{Castellani98a}. }.  
The question of identifying a material object, even if it is immediately accessible to the common sense, 
has always raised many philosophical issues once we consider it as the set of its 
replaceable constituents.   Heraclite's thoughts on the dynamical changes
and the paradox of talking about the ``same river'' while ``waters flows'' \cite[chap.~5, \S~2]{Graham08a}\nocite{Curd/Graham08a}
has never ceased to nurture  Western
philosophy (and Borges in particular). Another variation on these questions
goes back to an even more remote era when some Greek founding myths were forged: is it justified to talk about the ship of Theseus
after some  or even all of her original parts have been replaced?
However one can nevertheless suspect that all these issues can be reduced to a matter of
semantics (what is meant by ``same''), keeping in mind the danger of the inevitable tautology that plagues ontology while trying to
explain what ``existence'' signifies.

An attempt to give some empirical flesh to the question of indiscernability can be found in the writtings of Leibniz
who reports an observational test of the   \emph{principle of the identity of the indiscernibles} which is now attached to his name\footnote{plato.stanford.edu/entries/identity-indiscernible} \citep{Pesic00a}:

\begin{quotation}\small
\textsc{Philalethes.} A relative idea of the greatest importance is that of
identity or of diversity.  We never find, nor can we conceive it 'possible,
that two things of the same kind should exist in the same place at the same
time[. That is why, when] we demand, whether any thing be the same or
no, it refers always to something that existed such a time in such a place;
from whence it follows, that one thing cannot have two beginnings of
existence, nor two things one beginning...in time and place'.

\textsc{Theophilus.} In addition to the difference of time or of place there must
always be an internal \emph{principle of distinction}: although there can be many
things of the same kind, it is still the case that none of them are ever exactly
alike. Thus, although time and place (i.e. the relations to what lies outside)
do distinguish for us things which we could not easily tell apart by
reference to themselves alone, things are nevertheless distinguishable in
themselves. [\dots]

If two
individuals were perfectly similar and equal and, in short, \emph{indistinguishable}
in themselves, there would be no principle of individuation. I would even
venture to say that in such a case there would be no individual distinctness,
no separate individuals. That is why the notion of atoms is chimerical and
arises only from men's incomplete conceptions. For if there were atoms,
i.e. perfectly hard and perfectly unalterable bodies which were incapable
of internal change and could differ from one another only in size and in
shape, it is obvious that since they could have the same size and shape they
would then be indistinguishable in themselves and discernible only by
means of external denominations with no internal foundation; which is
contrary to the greatest principles of reason. In fact, however, every body
is changeable and indeed is actually changing all the time, so that it differs
in itself from every other. I remember a great princess, of lofty
intelligence, saying one day while walking in her garden that she did not
believe there were two leaves perfectly alike. A clever gentleman who was
walking with her believed that it would be easy to find some, but search
as he might he became convinced by his own eyes that a difference could
always be found. One can see from these considerations, which have until
now been overlooked, how far people have strayed in philosophy from the
most natural notions, and at what a distance from the great principles of
true metaphysics they have come to be.
\citet{Leibniz96a}
\end{quotation}
\begin{figure}[!ht]
\begin{center}
\includegraphics[width=.7\textwidth]{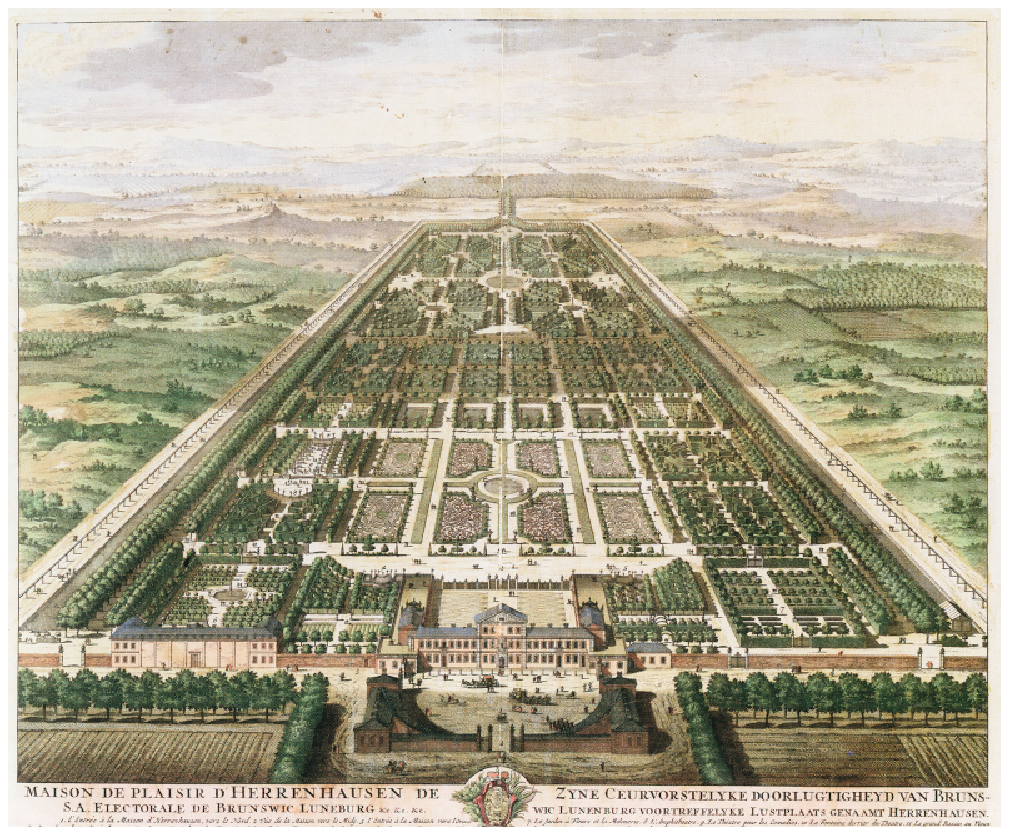}\includegraphics[width=.3\textwidth]{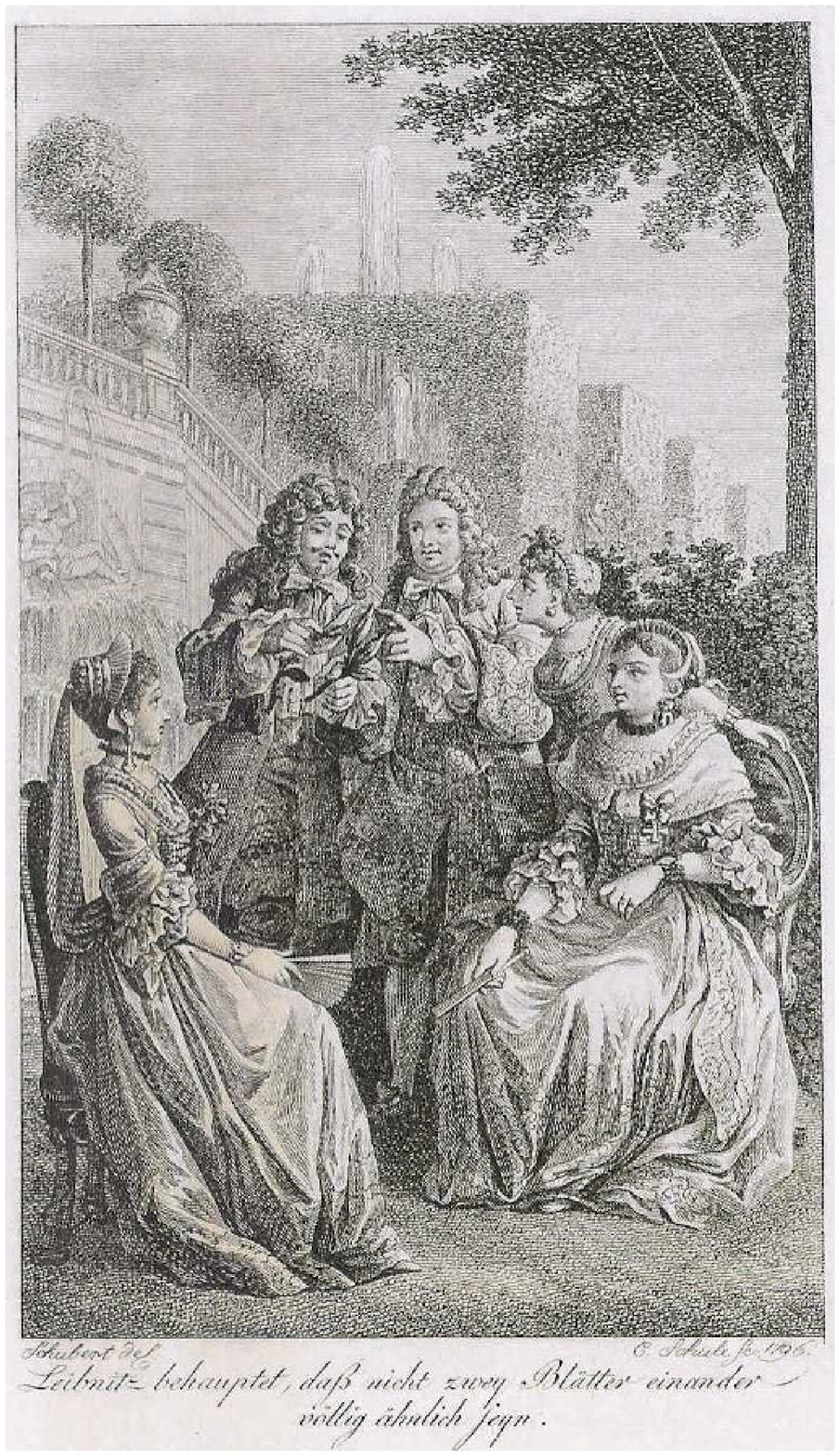}
\caption{\label{fig:leibniz} On the left: The Herrenh\"ausen Gardens around 1708 (wikipedia from Gottfried-Wilhelm-Leibniz-Bibliothek, Kartensammlung C Mappe 18 Nr. 178 b). On the right: Leibniz with Duchess Sophie, Carl August von Alvensleben and two ladies-in-waiting in the Herrenh\"ausen Gardens
(wikipedia from a biography of Leibniz by Johann August Eberhard published in 1795). }
\end{center}
\end{figure}

We learn from a letter written by Leibniz to  Sophie, Electress of Hanover (dated october, 31rd, 1705),
that the challenge took place in the Herrenh\"ausen gardens of Hanover  between princess Sophie and M . Carl August von Alvensleben,
the ``clever gentleman''. Sure the conclusion of the ``experiment'' would have been less straightforward
if instead of leaves, the bet had concerned bees
or ants (since clones are ubiquitous in one hive or in one anthill). However, in the same letter, maybe remembering the
geometric patterns of the garden itself (Fig.~1), Leibniz writes

\begin{quotation}\small
There are actual varieties everywhere and never a perfect uniformity in anything,
nor two pieces of matter completely similar to each other, in the great
as in the small. [\dots] Therefore there is always actual division and variation
in the masses of existing bodies, however small we go.
Perfect uniformity and continuity exists only in ideal or abstract things, as are
time, space, and lines, and other mathematical beings in which the divisions are not conceived as all done,
but as indeterminate and still feasible in an infinity of ways.
\citet[\S\,68. pp.~327--328, notes 669, 670]{Leibniz11a}
\end{quotation}

Retrospectively, and somehow ironically,
the above argument where perspires Leibniz' aversion against atomism  was founded. The existence of atoms not only do undermine  the
\emph{principle the identity of the indiscernibles} but the refutation is much stronger than Leibniz could have thought:
even the difference in position between atoms---a classical notion, on which, as Leibniz explains it, one may always rely to make a
distinction between similar material objects---becomes irrelevant at the quantum level as long as it is not measured. 

\section{Quantum abandon of individuality}

After a long maturation, mainly done in the first quarter of \textsc{xx}th century \citep{Spalek20a, Saunders20a}, starting with Planck's 1900
work on the blackbody radiation that can be
seen as the foundation stone of quantum physics, our concept of indistinguishability of
quantum particles
proceeds from the quantum theory of fields. Quantum particles, whether considered as elementary or composite, appears
as elementary excitations with respect to a ground reference state (the vacuum of the considered particles)
that are characterised by a handful of well-determined values which are the only observable quantities
that can be attributed \emph{simultaneously} to each of them: the mass, the electric charge, the spin and few other ``flavours''.
For instance, an electron is the particle whose mass is~$9.109\dots\times10^{-31}\,\mathrm{kg}$, whose
charge is~$e=-1.602\dots\times10^{19}\,\mathrm{C}$, whose spin is~$1/2$, etc. Other individual observable quantities, eventhough they remain
reasonably stable because of some conservation laws, for instance the linear momentum,
may be affected by an individual measurement of a non compatible quantity (the two observables do not commute in some precise algebraic sense) including, notably, the position of the particle. In fact, the orthodox interpretation leads to quantum properties that cannot be attributed simultaneously without raising contradictions with observations. According to the famous Heisenberg's inequalities,
after an arbitrarily precise measurement of its momentum, it is not that
we do not know the position of the particle, it is just that it does not have a precise position at all. In other words, a
measurement (say, the component~$J_z$ of the angular momentum along one direction~$z$)
does not affect the value of the previously measured non compatible quantity (say the component~$J_x$ of the angular momentum along an
orthogonal direction~$x$), it completely erases its existence: once~$J_z$ is measured, one cannot attribute an even unknown value
of~$J_x$ to the particle anymore. Therefore any quantum particle cannot have any history nor accidental properties nor  contingent secondary qualities that would still allow to individualise it, including its position\footnote{Some experiments have succeeded
  the technical challenge of keeping for several weeks one particle sufficiently localised away from the others. However,
  on the other hand, the correlations of entangled pairs over long distances show that the individuality cannot be based on spatial
  separation. Connected to the subject of the present text,
  the quantum contribution to the Western philosophical  analytic-reductionnist/holistic-emergentist
  dialectics  concerning the relationships between the part and the whole is fascinating \citep[for instance]{Maudlin98a}.
}.

These completely
counter-intuitive properties reflect all the more the stangeness of the quantum world that the number of particles
is itself a quantity that maybe incompatible with other observables\footnote{Some other interpretations try to fix
  what happens to be, as a last resort, a question of interpretation of quantum probabilities: alike what occurs in the classical world,
  do the latter reflect a lack of information or not? But as far as I know, these de Broglie-Bohmian points of view \citep[for a particularly
    interesting plea]{Bricmont16a} concern  a
  fixed number of particles only (generally one) and do not venture in the quantum field arena where even elementary particles can be created or annihilated.}. There are quantum \emph{pure} states (that is on which we have the maximal possible amount of information we can conceivably  get)
where the number of its constituents cannot be attributed.

Even in the cases where the number~$N$ of the particles can be attributed and maintained constant because the available energy is not
sufficient to create or destroy some of them,  the particles of the same species (electrons, neutral alcaline atoms, etc.)
still cannot be numbered, even in principle, for such a numbering is nothing but the attribution of a
discriminative quantity (essentially based on a position in a given configuration). The consequences are considerable if one
wants to study the statistical properties of a set of a such identical particles. We all know that
the odds (and the gains) to win at a trifecta horse race are significantly different if we decide to take into account or not the
finishing order of the top three horses. In statistical physics, it is energy (not money) that is distributed according to the
odds of the configurations and the distinguishability of the particles has observable consequences even at the macroscopical level.
The organisation of nucleons in the nuclei, of the electrons on the atoms which explain the chemical Mendeleiev
classification as well as the stability of matter, could not be explained if the particles were distinguishable.
The conducting properties of materials, notably the superconducting ones,
their thermal response, the superfluidity,  the behaviour of photons in a laser beam, or the existence
of states of matter like a Bose-Einstein condensate are emergent properties coming out from purely collective effect of some
set of particles where only the number of  its constituant has a physical meaning.

In fact, all quantum particles we know up to now
fall into two families according to their collective behaviour. The fermions, whose spin is half an integer,  design
particles that cannot share two identical states (the Pauli exclusion principle) whereas the bosons, of integer spin,
can condensate into the same individual state. All
the particles that constitute ordinary matter that we consider to be fundamental 
are fermions (mainly quarks, electrons). An assemblage of an odd number of fermions  remains a fermion whereas
any particle made of an even number of fermions (a Cooper pair of electrons, an Helium 4 atom for instance) follows a bosonic statistics. 

One remarkable thing is that this dichotomy can be understood with some topological arguments and the following tries to
give some hints about how this works. As a bonus, I will try to explain that for models in condensed matter where we can consider
that the dynamics lies in a layer whose effective dimension is~$D=2$, the same topological arguments offers more possibilities.
In two dimensions, one may consider some identical particles whose statistical behaviour is characterised by a continuous
parameter~$\theta$ that allows to interpolate continuously between bosons (say for~$\theta=0$) to fermions (say for~$\theta=\uppi$).
These existence of these \emph{anyons} (any-ons)---the word was coined by Frank \citet{Wilczek82a}---
has been proposed theoretically by Jon Magne Leinaas and Jan Myrheim (\citeyear{Leinaas/Myrheim77a}) forty-five years ago but it is only last year
that they received a first experimental
confirmation \citep{Bartolomei+20a}.

\section{Superpositions, interferences and phases}\label{sec:phase}

\begin{figure}[!ht]
\begin{center}
\includegraphics[width=\textwidth]{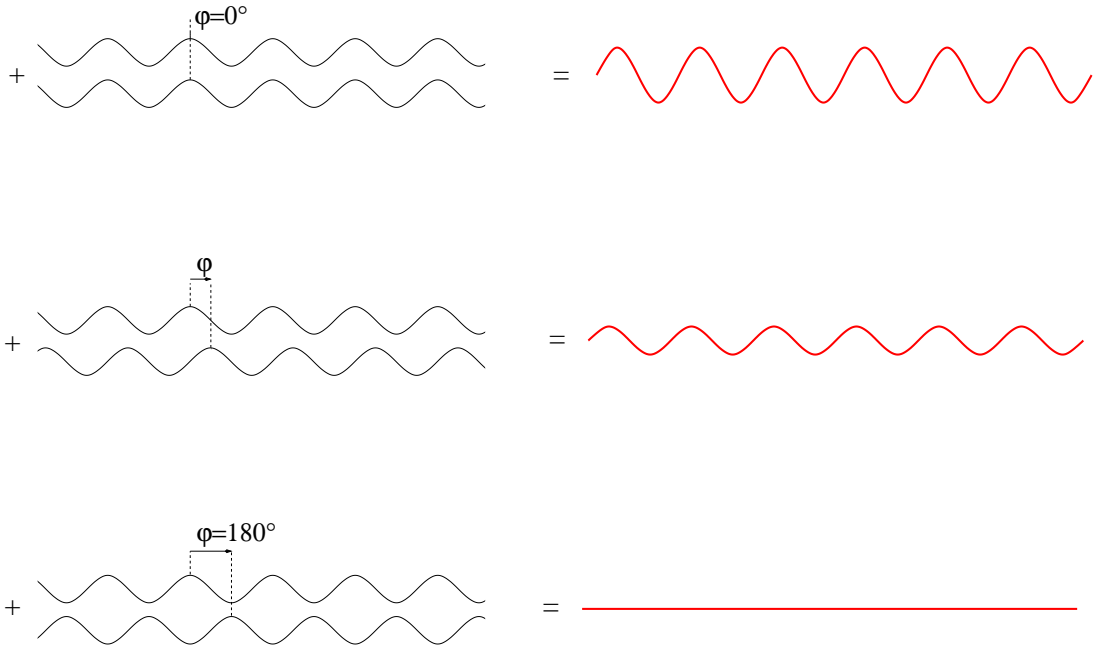}
\caption{\label{fig:phase} The resulting superposition of two waves with the same amplitude and frequency~$1/T$ is governed by
  their relative phase~$\upvarphi$. When there is no dephasing (upper case with~$\upvarphi=0$), two maxima (or minima)
  of the superposed waves coincide and they add constructively into a wave of maximal amplitude. Conversely, when the two waves have opposed phase
  (lower case with~$\upvarphi$ is half a turn), each ``bump'' is compensated by a ``hollow''; the two cancel one with the other
  and the resulting amplitude is almost zero. The relative phase is very much like an angle defined modulo one turn (or $360^\circ$),
  if the horizontal axis stands for the time, and~$\Delta t$ the time delay between two maxima, then~$\upvarphi=(\Delta t/T)\, 360^\circ$
  and one cannot distinguish between two delays differeing by an integer multiple of~$T$. To follow Dirac's argument in
  mathematical terms, the square modulus of the sum of
  two complex numbers~$z_1=|z_1|\text{\large e}^{\mathrm{i}\upvarphi_1}$ and~$z_2=|z_2|\text{\large e}^{\mathrm{i}\upvarphi_2}$ differs from~$|z_1|^2+|z_2|^2$
  because of the relative phase~$\upvarphi=\upvarphi_1-\upvarphi_2$: $|z_1+z_2|^2=|z_1|^2+|z_2|^2+2|z_1||z_2|\cos\upvarphi$.
The last term in the right-hand side is precisely the interference term.}
\end{center}
\end{figure}

To understand better, the reason why we must give up the systematic attribution of some properties to quantum objects
is that their pure states are described in terms of a (linear) superposition of states having definite properties. There is an experimentally accessible manifestation
of these superpositions: the interferences they can produce. Think of the most important
Young experiment on light diffracted by two holes on an opaque screen
that produces an interference pattern. Keeping a pure wave interpretation, the interference pattern is due to the superposition
of two waves, each one being diffracted by one hole (the other being closed). But once the two holes are opened
it is meaningless to say that the resulting wave has passed though one hole rather than the other.  Rather than getting a fuzzy or an unknown path, we actually completely loose the possibility of  attribution
of a path. These Young-like configuration, as well as others interference
experiments, have been
set up for individual quantum particles (photons but also electrons, atoms and even organic molecule made of hundred of atoms).
One crucial quantity that is measured in all these interference experiments is the \emph{relative phase}~$\upvarphi$ between the states being superposed, that is essentially an angle
given by the time delay between two periodic oscillations expressed in unit of their common period
(like an angle defined modulo one turn, only a delay modulo a period unit can be measured, see fig~\ref{fig:phase}).

These relative phases have been identified by Chen Ning Yang as one of the three melodies of theoretical physics
in the xxth century \citep{Yang03b}  and Dirac, looking back on the development of quantum physics that he contributed to shape,
writes
\begin{quotation}\small
  The question arises whether the noncommutation is really the main
  new idea of quantum mechanics. Previously I always thought it was
  but recently I have begun to doubt it and to think that maybe from
  the physical point of view, the noncommutation is not the only
  important idea and there is perhaps some deeper idea, some deeper
  change in our ordinary concepts which is brought by quantum
  mechanics. [\dots]
  [Following Heisenberg and Schr\"odinger], the
  probabilities which we have in atomic theory appear as the square of
  the modulus of some [complex] number which is a fundamental quantity. [\dots]
  I believe that \emph{this concept of probability amplitude is
    perhaps the most fundamental concept of quantum theory.}\par
  [\dots] The immediate effect of the existence of these probability amplitudes
  is to give rise to interference phenomena. If some process can take
  place in various ways, by various channels, as people say, what we
  must do is to calculate the probability amplitude for each of these
  channels. Then add all the probability amplitudes, and only after we
  have done this addition do we form the square of the modulus and get
  the total result for the probability of this process taking
  place. You see that the result is quite different from what we
  should have if we had taken the square of the modulus of the
  individual terms referring to various channels. It is this
  difference which gives rise to the phenomenon of interference, which
  is all pervading in the atomic world [\dots].\\  
So if one asks what is the main feature
    of quantum mechanics, I feel inclined now to say that it is not
    noncommutative algebra. It is the existence of probability
    amplitudes which underlie all atomic processes. Now a probability 
   amplitude is related to experiment but only partially. The
    square of its modulus is something that we can observe. That is
    the probability which the experimental people get. But besides
    that there is a phase, a number of modulus unity which can modify
    without affecting the square of the modulus.  And this phase is
    all important because it is the source of all interference
    phenomena but its physical significance is obscure. So the real
    genius of Heisenberg and Schr{\"o}dinger, you might say, was to
    discover the existence of probability amplitudes containing this
    phase quantity which is very well hidden in nature and it is
    because it was so well hidden that people hadn't thought of
    quantum mechanics much earlier. \cite[pp.~154--158]{Dirac72a}
\end{quotation}

\section{Feynman's paths}

\begin{figure}[!ht]
\begin{center}
\includegraphics[width=.5\textwidth]{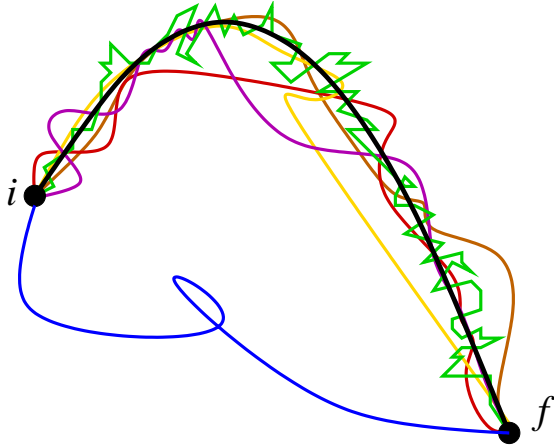}
\caption{\label{fig:feynmanparabole}For one particle starting at initial position~$i$, its probability amplitude to find it in the final position~$f$ is given by the sum~\eqref{eq:Zif} over all histories
  i.e. all the continuous (non necessarily differentiable) possible trajectories connecting~$i$ to~$f$. Only a bunch of trajectories
  in a neighbourhood---whose size is governed by Planck's constant--- of the classical Newtonian trajectory
  (here the black parabola for a particle in a uniform constant force field) will contribute with a constructive interference (the phases
  are proportional to the classical action which is stationary).
  Any other bunch of trajectories far from the classical one (say around the blue lower trajectory) will bring a negligible contribution because their
  phases~$\upvarphi_h$ vary extremely rapidly and provoque a destructive interference.}
\end{center}
\end{figure}

Half a century after Dirac showed the equivalence of  Schrodinger's ``wave mechanics''
and the Heisenberg's ``matrix mechanics'' in a unified formalism, Feynman proposed in his thesis of 1942 \citep{Feynman42a} a third way of
computing quantum predictions. Also equivalent to the first ones, Feynman's formalism, at the price of introducing a
subtly new mathematical concept of functional integration, gives to the quantum interferences the first role\footnote{ The Dirac's quotation given at the end of section~\ref{sec:phase} is the transcription
  of a conference he gave  in April 1970 for a general audience.
  Though he talks about the recent development in quantum electrodynamics, and though he defends the idea that
\emph{this concept of probability amplitude is
    perhaps the most fundamental concept of quantum theory}, surprisingly enough,
Dirac does not mention Feynman at all in his text. The only very vague allusion I could find lies, perhaps, in the sentence immediately following the quotation above: \emph{If you go
        over the present day theory to see what people are doing you
        find that they are retaining this idea of probability
        amplitude} \cite[p.~158]{Dirac72a}.  }: the probability amplitude~$Z_{i\to f}$
for a system to evolves from an initial state~$i$ to a final state~$f$
are explicitely written as the result of the interference between all the possible histories the system may follow between the two
states: up to a normalisation factor we can write\footnote{The writing is simplified (it hides that
  the sum  covers an infinite functional continuum)
  but captures the spirit of the famous
  Feynman path-integrals.}
\begin{equation}\label{eq:Zif}
  Z_{i\to f}= \sum_{\substack{\text{all possible histories~$h$}\\\text{connecting $i$ to~$f$}}}|z_h|\,\text{\large e}^{\mathrm{i}\upvarphi_h} 
\end{equation}
Each virtual history~$h$ is weightened by a complex number~$z_h$ whose phase is~$\upvarphi_h$.
Only for a subset of histories that interfer constructively, 
we can think of a quasi-classical evolution--- but still fuzzy at Planck's constant scales.
Most of virtual histories, even when they have the same amplitude~$|z_h|$,  have a phase
that differ too much from the average of the classical bunch and their contribution is mostly destroyed by their
neighbours (Fig.~\ref{fig:feynmanparabole}).
One possible history of one particle is just given by one continuous path made of all possible positions in ordinary space
(its trajectory) and the constructive interference occurs precisely
when it satisfies Newton's (or Euler-Lagrange's, or Hamilton's) classical equations.

However as soon as two or more particles evolve, even if non interacting, the appropriate place to describe
histories is not the ordinary space anymore but a more abstract space, the \emph{configuration space}
(the same space on which the Schr\"odinger wavefunctions are defined). To simplify the discussion,
we will assume that the initial state has been prepared with a determined number~$N$ of particles of the same species
and this number will be maintained all along the evolution.
In that case\footnote{When particles can be created or annihilated, one cannot avoid working with fields whose configuration space
  is infinitely dimensional. This is almost unavoidable when dealing with quantum electrodynamics since photons
  are massless particles and thus can have an  arbitrary low individual energy; therefore are cheap to create and easy to absorb.
  It requires
a tremendous experimental skill to preserve a fixed number of photons for a while (in a superconducting cavity for instance).},
the configuration space has~$ND$ dimensions where~$D$ is the dimension of ordinary space (most of the time, obviously, it is 3
but in some condensed matter models, we shall see that~$D$ can be lowered to 2 or even~1).
To try to visualise the evolution of such a system,
one may come back to the ordinary~$D$-dimensional space where inevitably each particle can be individualised
by a numbering  that is continuously followed as they evolve from a given initial state to a given final state.
But as we explained in the previous section, such a numbering has no physical basis at the quantum level
and any continuous permutation among them not only can but must be considered among the Feynman's histories.
It is also crucial to keep in mind in the following that the inevitable interactions between particles (the photon being aside, see the last footnote) exclude the histories where two of them overlap (for the corresponding energy of such a configuration diverges which make
the phase oscillating infinitely quickly which destroy the superposition).

\section{Where the topological properties come from}

It is to \citet{Dirac31a} that we owe the first identification of
a quantum property of topological origin, namely a universal
constraint on the electric charges if there would exist magnetic
monopole\footnote{Monopoles would  come with a violation of  the Maxwell equation
  $\mathrm{div}B=0$.}.  By Feynman's own admission,
\citet{Dirac33a} was also an inspiring source for the implementation of
the ideas of the sum over histories briefly sketched in the previous
section.  Indeed, the sum over histories in~\eqref{eq:Zif} probes
directly the topology of the space where the histories take place,
namely the configuration space. In particular, since the histories are
necessarily continuous they can be classified according to the
so-called first homotopy group of the configuration space, each
homotopy class being made of all the histories that can be
continuously deformed one into another.

For one particle in ordinary space, most often, the homotopy
properties are trivial in the sense that only one class generally
exists. Like in Fig.~\ref{fig:feynmanparabole}, all the paths can be
continously deformed one into another\footnote{One can manufacture on
  purpose some holes in space by creating zones that are forbidden to
  the paths by a magnetic field.  This is the celebrated geometry
  proposed by Ehrenberg-Siday-Aharonov-Bohm
  \citep{Ehrenberg/Siday49a,Aharonov/Bohm59a,Aharonov/Bohm61a} where
  quantum topological phases are involved even for just one
  particle.}. But as soon as at least two identical particles are
involved more than one homotopy class should be considered.

Whenever several homotopy classes are present, some new possibilities are offered in the Feynman's
approach~\cite[and its references for a rigourous justification]{Mouchet20a}.
Because of the well-behaved composition law of histories together with some conservation of probabilities, one can
attribute to each class~$\mathfrak{c}$ a phase~$\chi_{\mathfrak{c}}$ that depends only on~$\mathfrak{c}$
and not on any specific choice of one of its member.
This is why they are called \emph{topological phases} and
then~\eqref{eq:Zif} can be extended to  
\begin{equation}\label{eq:Zifchi}
  Z_{i\to f}= \sum_{\substack{\text{all homotopy classes~$\mathfrak{c}$}\\ \text{of histories connecting $i$ to~$f$}}}
    \text{\large e}^{\mathrm{i}\chi_\mathfrak{c}} \sum_{\substack{\text{all histories~$h$ in~$\mathfrak{c}$}}}|z_h|\,\text{\large e}^{\mathrm{i}\upvarphi_h}\;.
\end{equation}
Of course when only one class is present or if we take~$\chi_{\mathfrak{c}}=0$ for all classes one recovers~\eqref{eq:Zif} but it happens
that other choices are actually realised. 

\section{Permutations and braids}

\begin{figure}[!ht]
\begin{center}
\includegraphics[width=.7\textwidth]{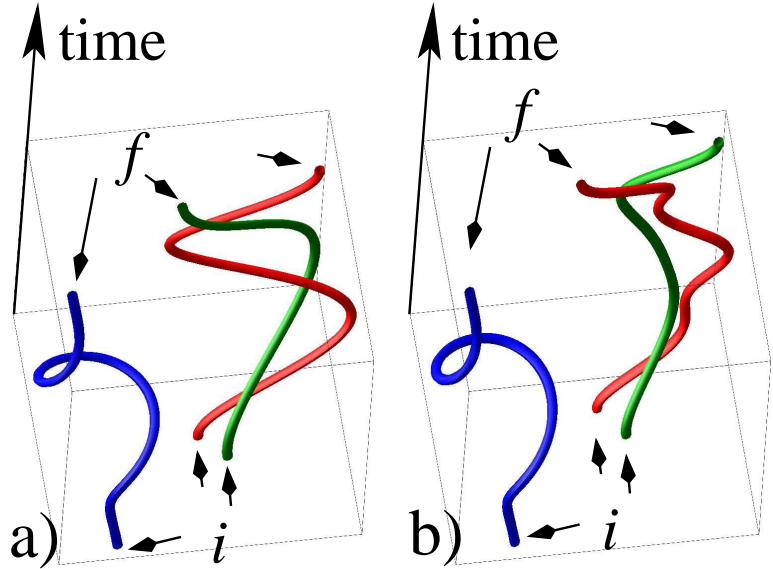}
\caption{\label{fig:tressetriple} Representation of two Feynman histories (or paths in the configuration space)
  for~$N=3$ particles. Time axis is chosen to be
  vertical while ordinary~$D$-dimensional space is perpendicular to it.
  Since on these examples, the red and green threads do not connect the smae initial and final points,
  these histories cannot be continously deformed one into the other and therefore
  paths a) and b) belongs to two different homotopy classes. }
\end{center}
\end{figure}

Consider for instance~$N=3$ identical particles. Figure~\ref{fig:tressetriple} provides two examples of histories represented in ordinary
space with the time being given by the vertical axis.

\begin{figure}[!ht]
\begin{center}
\includegraphics[width=\textwidth]{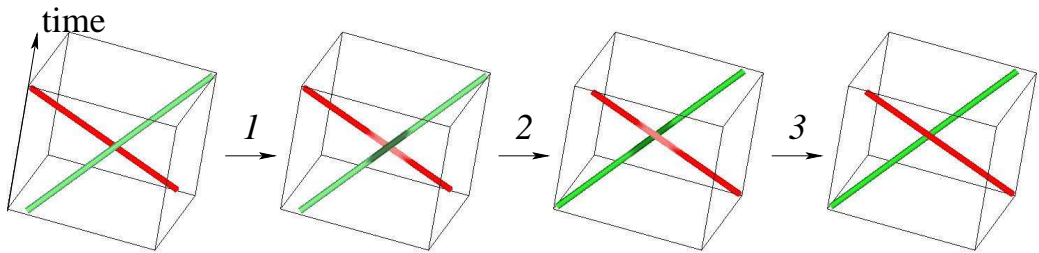}
\caption{\label{fig:croisement} In these graphical representations of a path for~$N=2$ particles, the time axis is plotted vertically
  while the ordinary space is perpendicular. When~$D=3$, moving in the third spatial dimension can be thought as changing the darkness of the threads.
  Because no particle can be at the same place at the same time,   in~$D=2$ the threads cannot cross. In~$D=3$,
  one can always continuously separate them in the third dimension without any cut:
  in step~1 a darkening of a portion of one thread (here the green one) and a lightening of a portion of the other (the red one).
  Then, a crossing of these two portions in  the two other dimensions is possible and, eventually,
  in step~3 one can restore the initial value of the third spatial coordinate. }
\end{center}
\end{figure}

\begin{figure}[!ht]
\begin{center}
\includegraphics[width=.25\textwidth]{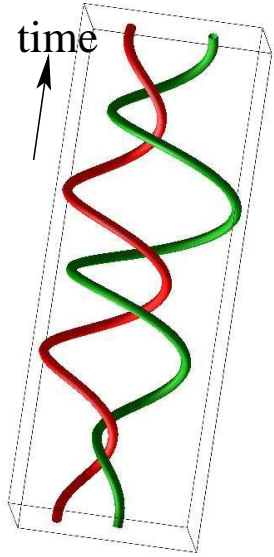}
\caption{\label{fig:tressedouble} In~$D=3$, the crossing of two threads being possible, one can always entangle
  a succession of an arbitrary number of exchanges of two particles, leaving us with two homotopy classes only
  (the identity and the transposition).
  In~$D=2$, the braid such obtained
is always different from the previous ones and the homotopy classes can be labeled by a unbounded integer.}
\end{center}
\end{figure}

As explained above, because the particles cannot overlap, we can follow each individual trajectory by
a thread that allows to keep their individualisation (by a numbering or, more visually, a color).
In the two examples drawn, the two histories differ by a permutation of particles: in the final states the end
position of the ``green'' thread has been exchanged with the end position of the ``red'' one. And precisely because of
this permutation, one cannot deform one history into another without cutting two threads before gluing the pieces
appropriately
which would break the continuity.
The representation chosen here can involve two spatial dimensions only (the horizontal plane). When~$D=3$ one can imagine
that the motion in the third spatial dimension is represented by a change of darkness in the color of the thread. With this image
in mind, one can understand that when~$D=3$ one can have the thread crosses one with an other even though two
particles still cannot be at the same place at the same moment (Fig.~\ref{fig:croisement}). This latter constraint forbids two threads
to cross at some point where their darkness is the same (all the three coordinates would coincide)
but one can always bypass this restriction by changing the
the darkness of one thread at a point of crossing (that is moving it in the third spatial dimension) then
cross the threads at this point and restore the original darkness after this operation.
In summary, for~$D=3$, when using the graphical representation of an history given in figure~\ref{fig:tressetriple} or~\ref{fig:tressedouble}, the thread can be crossed
but not in~$D=2$.
This latter statement makes all the topological difference between~$D=2$ and~$D=3$. In the latter case, on can see that
each homotopy class is in fact an ordinary permutation, whereas the structure of the classes in~$D=2$ is much richer.
By concatenating the threads in order to represent the composition of two evolutions, we introdcuce naturally an algebraic
internal law that makes the set of classes a group (the so-called fondamental homotopy group of the configuration space).
Then, by a straightforward generalisation to~$N$ identical particles, in~$D=3$ the topological group is simply the \emph{permutation group}
of~$N$ elements whereas in~$D=2$, the group is called the \emph{braid group} of~$N$ strands.
Moreover, it can be shown that the topological phases must naturally compose accordingly: if~$\mathfrak{c}\cdot\mathfrak{c}'$ stands for the class obtained
by concatenating the~$N$ threads in~$\mathfrak{c}$ with the~$N$ threads in~$\mathfrak{c}'$, then,
is can be shown that  we must have
\begin{equation}
   \text{\large e}^{\mathrm{i}\chi_{\mathfrak{c}\cdot\mathfrak{c}'}} = \text{\large e}^{\mathrm{i}\chi_\mathfrak{c}}  \text{\large e}^{\mathrm{i}\chi_{\mathfrak{c}'}}\;. 
  \end{equation}
For~$D=3$, the latter relation
leaves us with a simple alternative
since the~$(2p+1)$\th iteration of a
transposition of two threads, $p$ being an integer,
leaves us with the transposition
itself, that is~$\text{\large  e}^{\mathrm{i}(2p+1)\chi_{\mathfrak{c}}}=\text{\large  e}^{\mathrm{i}\chi_{\mathfrak{c}}}$
whose solution can only be
\begin{equation}
          \text{\large e}^{\mathrm{i}\chi_{\mathfrak{c}}}=1\qquad\mathrm{or}\qquad\text{\large e}^{\mathrm{i}\chi_{\mathfrak{c}}}=-1
        \end{equation}        
for all the classes~$\mathfrak{c}$ associated with a transposition of two identical particles. The first choice corresponds to bosons
and the second to the fermions.
        For~$D=2$, if~$\mathfrak{c}$ is a bread with two threads associated with a transposition
        one can iterate the concatenation of the same bread made of two threads an arbitrary number of times and the result is always a different braid
        because of the interdiction of crossing the threads (Fig.~\ref{fig:tressedouble}). The algebraic constraints on the bread group are much less restrictive that for the permutation group and in particular we can take consistently
        \begin{equation}
          \text{\large e}^{\mathrm{i}\chi_{\mathfrak{c}}}=\text{\large e}^{\mathrm{i}\theta}
          \end{equation}
where~$\theta$ is an angle,  that characterises the species of the identical particles under consideration: it continuously interpolates
between the bosons ($\theta=0$) and the fermions~($\theta=\uppi$). These is precisely this quantity that characterises an anyon.
This phase beeing relative to some terms in a quantum amplitude of probability, they have observational consequences through interferences. 

This is the way we can connect the statistical physical properties  of~$N$ identical quantum particles (the bosonic, fermionic or even anyonic character
under permutations) and some
topological properties (the first homotopy group of their configuration space).

\ifx\undefined\BySame
\newcommand{\BySame}{\leavevmode\rule[.5ex]{3em}{.5pt}\ }
\fi
\ifx\undefined\textsc
\newcommand{\textsc}[1]{{\sc #1}}
\fi
\ifx\undefined\emph
\newcommand{\emph}[1]{{\em #1\/}}
\fi

\end{document}